# Femtosecond-to-millisecond holographic imaging of laser ablation dynamics

Shotaro Kawano[1], Keiichiro Toda[1], Miu Tamamitsu[1], Haruyuki Sakurai[1], Kuniaki Konishi[1], and Takuro Ideguchi[1,2,3*]

[1]Institute for Photon Science and Technology, The University of Tokyo, Tokyo, Japan

[2]Department of Advanced Materials Science, The University of Tokyo, Tokyo, Japan

[3]RIKEN Center for Advanced Photonics, RIKEN, Saitama, Japan

[*]ideguchi@edu.k.u-tokyo.ac.jp

## Abstract

Femtosecond laser ablation redistributes optically deposited energy across electronic, structural, mechanical, and thermal degrees of freedom over timescales from femtoseconds to milliseconds. However, these coupled processes are usually measured in separate temporal ranges and through different observables, limiting quantitative comparison between early transient dynamics, residual heating, and final morphology. Here we introduce pump–probe holographic imaging that reconstructs amplitude- and phase-resolved optical fields across this full temporal range under matched imaging conditions. Applied to deep-ultraviolet femtosecond ablation of BK7 glass, the method captures the transition from early excitation and removal-stage dynamics to residual substrate heating and permanent modification. Differential phase analysis isolates sub-nanosecond evolution of the transient ablating layer and microsecond residual heating after material removal. Above the ablation threshold, crater depth increases with fluence, whereas the residual thermal signal saturates, indicating that additional absorbed energy is preferentially partitioned into material removal and ablation-related processes rather than retained as substrate heat. These results identify fluence-dependent energy partitioning as a dynamical basis of low-heat-affected femtosecond processing and establish holographic imaging as a route to tracking laser-driven nonequilibrium material transformations.

## Introduction

Femtosecond laser ablation has become a powerful tool for high-precision fabrication across a wide range of materials, including transparent dielectrics. Compared with conventional nanosecond laser processing, femtosecond excitation suppresses heat diffusion around the processed region, thereby reducing melt formation and recast effects (1). This low-heat-affected nature has been inferred from final processed morphologies, such as reduced heat-affected zones, yet its dynamical origin cannot be fully captured by the simple statement that heat diffusion is limited during an ultrashort pulse. After the initial electronic excitation, absorbed energy is redistributed through carrier relaxation, structural transformation, phase transitions, plasma formation, material ejection, and residual substrate heating. Understanding femtosecond ablation therefore requires tracking how optical energy flows through electronic, structural, mechanical, and thermal channels before determining the final processed morphology.

Previous studies have established the characteristic timescales of individual processes involved in femtosecond laser ablation. Carrier excitation occurs on the femtosecond timescale, followed by energy transfer to the lattice within picoseconds (2,3). Above the ablation threshold, material removal emerges on nanosecond timescales (4–6), while residual thermal energy dissipates into the surrounding material over microsecond-to-millisecond timescales (7). Pump–probe imaging has become a central approach for investigating these dynamics by tracking transient changes in the optical response of a time-delayed probe pulse. Mechanical delay scanning enables direct observation of ultrafast dynamics from femtoseconds to nanoseconds (4–6,8–20), whereas electronically controlled timing schemes extend the accessible delay range to much longer timescales (7,21). In principle, combining these approaches should allow laser-processing dynamics to be reconstructed across an exceptionally broad temporal window. In practice, however, early excitation, transient material removal, residual heating, and final morphology have often been measured in separate temporal windows or under different imaging conditions, limiting direct comparison across the full processing sequence.

A further limitation is that many previous pump–probe studies spanning multiple temporal regimes have visualized laser-processing dynamics primarily through intensity changes of the probe light, which mainly reflect attenuation, scattering, structural modification, or material removal (22,23). Intensity-only measurements do not provide phase-sensitive information on the transient optical properties of the material. Complex-field imaging based on digital holography addresses this limitation by reconstructing both the amplitude and phase of the optical field and has found broad applications ranging from materials characterization to label-free biomedical imaging (24,25). Phase imaging is particularly informative because it reflects refractive-index variations associated with transient carrier populations, stress fields, ablation plumes, structural modification, and thermal fields (4,7,9,10,15,16). However, complex-field measurements performed with different probe sources, fields of view, spatial sampling, or other experimental conditions are difficult to compare quantitatively across widely separated timescales. A unified complex-field imaging framework is therefore needed to follow energy flow from the initial optical excitation to the final residual thermal and structural state.

Here, we present a pump–probe holographic imaging platform for femtosecond-to-millisecond complex-field imaging of laser ablation dynamics. The system combines off-axis digital holography with wide-range delay control to reconstruct both transmittance and optical path difference (OPD) maps under matched imaging conditions. A common-path interferometric geometry provides phase sensitivity sufficient to detect weak transient thermal signals, while image registration onto a common pixel grid enables differential analysis between widely separated delay times. Using deep-ultraviolet femtosecond ablation of BK7 glass as a model system, we first reconstruct the full temporal evolution of complex-field signatures associated with carrier excitation, stress generation, structural changes, material removal, plume expansion, permanent modification, and thermal diffusion. We then use differential phase analysis to isolate two representative components of the energy-flow pathway: sub-nanosecond optical evolution of the transient ablating layer and microsecond residual substrate heating after ablation. The latter analysis reveals a fluence-dependent energy-partitioning behavior in which crater depth increases above the ablation threshold while residual substrate heating saturates. This observation shows that additional absorbed energy is preferentially redirected into material removal and ablation-related processes rather than retained as heat in the remaining substrate.

## Results

### Femtosecond-to-millisecond pump–probe holographic imaging platform

Figure 1a shows the pump–probe holographic imaging platform, including the light sources, delay-control scheme, and digital holographic imaging system. A femtosecond regenerative-amplifier laser system delivers 200-fs pulses at 1030 nm, which are divided into pump and probe arms. Deep-ultraviolet (DUV) pump pulses at 257 nm and visible probe pulses at 515 nm are generated through fourth- and second-harmonic generation, respectively. For delays below 3 ns, the pump–probe delay is scanned using a mechanical optical delay line. At longer delays, timing control is achieved electronically using a separately synchronized probe source. Specifically, a Yb-fiber laser generates 1-ns probe pulses at 532 nm through second-harmonic generation and is electronically synchronized to the pump laser. The temporal resolution is therefore determined by the probe duration, providing 200 fs resolution below 3 ns and 1 ns resolution at longer delays.

The pump and probe beams are combined coaxially and delivered to the sample. The pump beam is focused to a spot diameter of ~20 µm, whereas the probe beam provides collimated illumination over a field of view exceeding 300 µm (Fig. 1b). The pump-beam profile was characterized independently using a custom beam profiler (26) and remained stable throughout each measurement. For some datasets, the focusing condition was intentionally adjusted according to the measurement purpose, as described in Supplementary Note 1.

After transmission through the sample, the probe beam enters a common-path off-axis digital holography (DH) interferometer (Fig. 1c). A diffraction grating separates the probe beam into zeroth- and first-order diffracted beams, which serve as the object and reference waves, respectively. A 4f imaging system relays the object wave containing the laser-processed region onto the image sensor. The reference wave is laterally sheared after sample transmission and overlapped on the sensor with the object wave, such that the laser-processed region in the object wave interferes with a corresponding unprocessed region in the reference wave. This interference generates an off-axis hologram (Fig. 1d), from which maps of intensity transmittance, hereafter referred to as transmittance, and OPD are reconstructed by Fourier analysis (Fig. 1e). Additional experimental details are provided in Methods and Supplementary Note 2.

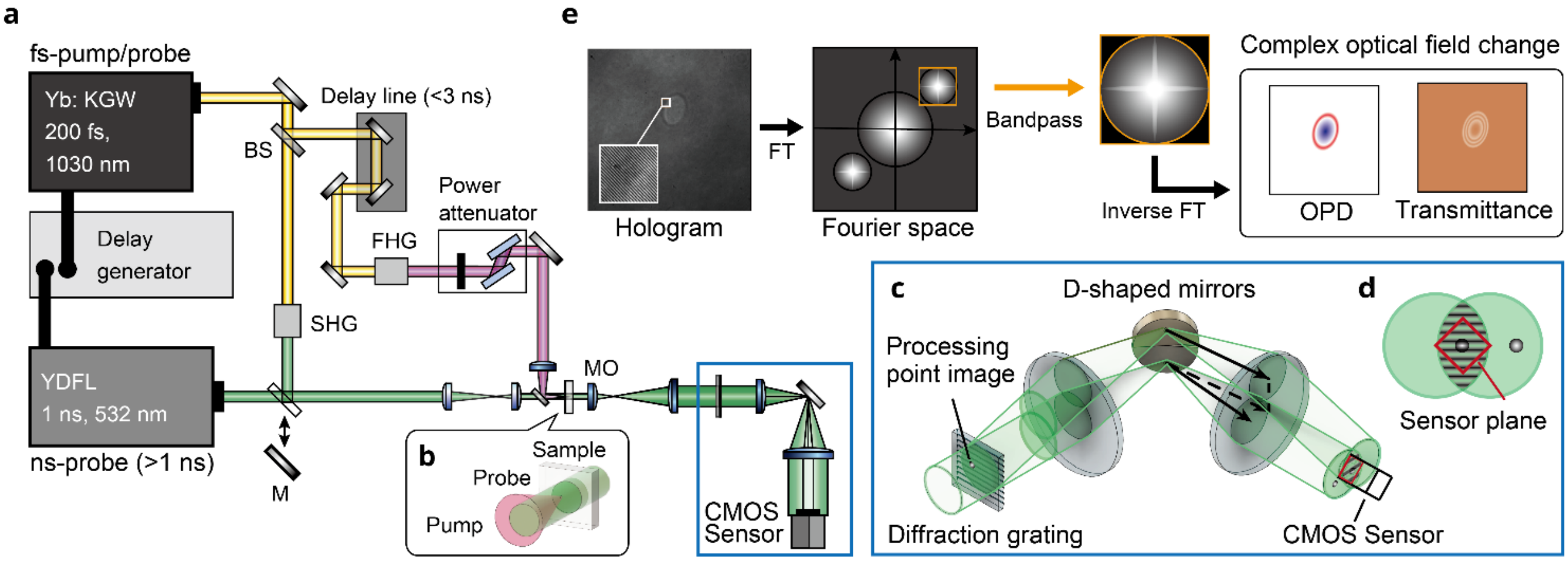

**Figure 1 | Femtosecond-to-millisecond pump–probe holographic imaging platform. a,** Optical layout of the pump–probe holographic imaging system, including the femtosecond pump and probe sources, nanosecond probe source, delay-control scheme, and digital holographic imaging path. BS, beam splitter; M, flip-mounted mirror; MO, microscope objective. **b,** Spatial overlap of the focused pump beam and collimated probe beam on the sample surface. The pump beam excites the laser-processing region, whereas the probe beam illuminates a wider field of view for holographic imaging. **c,** Diffraction-grating-based common-path shearing interferometer used for off-axis digital holography. The object and reference waves are formed by splitting the probe wave after sample transmission into two diffracted beams that are laterally sheared on the image sensor. **d,** Representative off-axis shearing interferogram recorded on the image sensor. **e,** Reconstruction workflow for retrieving transmittance and OPD maps from the recorded hologram. FT, Fourier transform.

**Phase sensitivity and spatial resolution**

We evaluated the noise performance and spatial resolution of the holographic imaging system. To characterize temporal fluctuations, 101 complex-field images were recorded at 30 Hz without pump excitation, and 100 consecutive differential images were analyzed. The resulting single-frame noise floors were 2–5 mrad for OPD and 0.4–2.0% for transmittance (Supplementary Note 3). This phase stability is sufficient to resolve weak transient thermal signals in femtosecond laser processing, which are typically on the order of 10 mrad. The transmittance noise is also sufficient to resolve the percent-level attenuation and scattering signals observed during carrier excitation and material removal. The present system therefore provides improved phase sensitivity compared with earlier time-resolved interferometric imaging studies, which reported representative values of ~30 mrad phase noise and ~1% intensity fluctuation (9).

Spatial resolution was evaluated using a test target. After aberration compensation and digital refocusing, the system achieved a near-diffraction-limited spatial resolution of ~620 nm (Methods and Supplementary Notes 4 and 5). This resolution is sufficient to resolve micrometer-scale crater morphologies and is comparable to the best values reported previously for time-resolved laser-processing imaging (15).

**Complex-field visualization of laser ablation dynamics from femtoseconds to milliseconds**

Using the pump–probe holographic imaging system, we reconstructed the temporal evolution of the complex optical-field response during femtosecond laser ablation from the initial optical excitation to the final post-processed state. Experiments were performed on BK7 glass at peak pump fluences of 0.9 and 1.7 J cm$^{-2}$, slightly below and above the ablation threshold of 1.4 J cm$^{-2}$, respectively (Supplementary Note 6). BK7 glass has a bandgap of 3.8 eV (27) and linearly absorbs the 4.8-eV DUV pump photons. Figure 2a summarizes the characteristic timescales and physical mechanisms established in previous studies (2–8,10–15,20,22,28–32). Figures 2b and 2c show the corresponding time-resolved transmittance and OPD maps measured using the present system. Comparison with the known temporal hierarchy of femtosecond laser processing enables assignment of the observed complex-field signatures to distinct stages of the energy-flow pathway.

We first examined the sub-picosecond regime, where the responses below and above the ablation threshold were qualitatively similar. During the pump pulse irradiation, from −0.13 to 0.13 ps delay, both transmittance and OPD decreased within the excitation region. Temporal traces extracted at the beam center showed nearly monotonic reductions in both observables (Fig. 2d, gray region), consistent with rapid free-carrier generation. The dense carrier population both attenuates probe transmission and lowers the refractive index through the free-carrier contribution (2,3,8,10,12–15,20). Superimposed on this dominant negative response, a weak positive OPD signal appeared near the edge of the excitation region (Fig. 2e, inside the black dotted line), likely reflecting a photoelastic response induced by carrier-generated stress (20).

Although the overall sub-picosecond dynamics resemble previous observations of femtosecond laser excitation in dielectrics driven by multiphoton absorption (8,10,12,14,15), two notable differences were observed. First, earlier studies based on multiphoton excitation frequently reported a pronounced positive OPD contribution associated with the optical Kerr effect during carrier excitation. In contrast, the present measurements exhibited no comparable Kerr-dominated response. This difference is consistent with the linear-absorption-driven excitation pathway in BK7 glass under DUV irradiation, which suppresses the relative contribution of the nonlinear Kerr response. Second, the OPD decrease observed above the ablation threshold was smaller than that observed below threshold (Fig. 2d, gray region). This behavior suggests accelerated carrier relaxation at high excitation density, potentially associated with stronger carrier localization (Supplementary Note 7) and enhanced carrier scattering near the surface region (Supplementary Note 8).

In the picosecond-to-sub-nanosecond regime, the OPD distributions evolved substantially, whereas the transmittance maps remained qualitatively similar to those in the sub-picosecond regime. At the beam center, the OPD rapidly recovered between 1 and 10 ps (green region in Fig. 2d), followed by a slower increase extending from 10 to 500 ps (orange region in Fig. 2d). The initial recovery is attributed primarily to carrier trapping (8,10,12–15), whereas the subsequent slower evolution likely reflects overlapping contributions from transient lattice heating and structural modification associated with defect formation and amorphous-network reorganization (31,32). Above the ablation threshold, the onset of the phase transition leading to material removal is also expected within this temporal window (22). Consequently, the observed OPD evolution in the above-threshold regime contains overlapping contributions from thermal, structural, and ablation-related dynamics that cannot be directly separated within this timescale alone. We later exploit differences in their temporal evolution to isolate these contributions through differential analysis.

Distinct spatial features also emerged near the edge of the excitation region. At 10 ps delay, the OPD maps exhibited negative peripheral signals surrounded by weak positive variations (between the black and white dashed lines in Fig. 2f). This pattern is consistent with overlapping contributions from residual free carriers (negative OPD) and carrier-induced stress (positive OPD). Previous studies showed that stress-related signals decay rapidly through the generation of compressive stress waves propagating into the surrounding material (13,20). By contrast, the negative peripheral OPD associated with long-lived free carriers persisted up to ~1 ns (blue arrows in Figs. 2b and 2c).

The dynamics diverged substantially in the nanosecond regime as material removal emerged above the ablation threshold. At 10 ns delay, the sub-threshold OPD map exhibited a broad positive distribution, whereas the above-threshold map showed a central depression accompanied by surrounding circular structures (blue dotted circles in Fig. 2c). These features are attributed to crater formation and transient ablation-plume expansion (4,5,28). The central negative OPD arises from replacement of the crater region by ejected material and air following surface removal. Because air has a lower refractive index than BK7 glass, this replacement produces a negative OPD contribution. Meanwhile, the circular structure evolved into an expanding ring-shaped positive OPD between 8 and 20 ns (Fig. 2g). This feature is attributed to refractive-index enhancement in the compressed gas behind the shock front generated by plume expansion (4). Consistent with this interpretation, the measured wavefront trajectory (blue dashed line in Fig. 2g) follows the expected scaling behavior of an acoustic shock wave propagating in air (Supplementary Note 9).

In the microsecond regime, both fluence conditions exhibited positive OPD variations, while the above-threshold measurements retained a central depression associated with the crater region. At ~1 μs delay, the positive OPD distribution outside the crater consisted of a transient component and a persistent component. The transient contribution decayed with a characteristic timescale of ~100 μs (red region in Fig. 2d), consistent with thermal diffusion in BK7 glass estimated from the heated volume size and thermal diffusivity (Supplementary Note 10). After relaxation of this thermal component, a residual positive OPD remained in the final state (black arrow in Fig. 2f), indicating permanent Type-I modification (30–32). In BK7 glass, this persistent refractive-index increase is attributed primarily to defect-related structural reorganization rather than the densification mechanisms commonly discussed for fused silica (30,31).

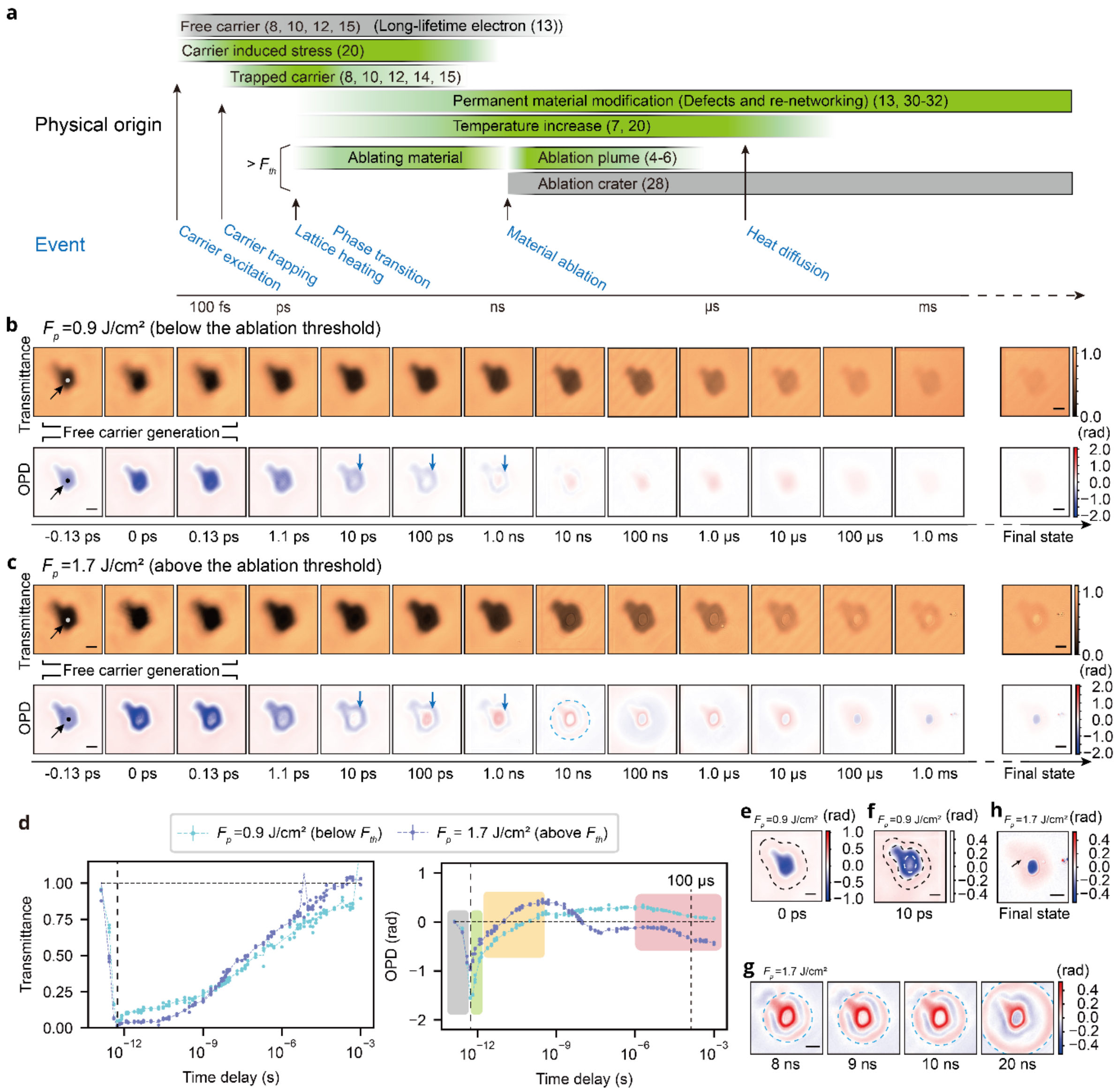

**Figure 2 | Complex-field visualization of femtosecond laser ablation from femtoseconds to milliseconds. a,** Timeline of characteristic processes involved in femtosecond laser ablation and their corresponding signatures in the complex optical-field response. Signal contributions predominantly observed in the OPD channel are highlighted in green, whereas those visible in both OPD and transmittance are shown in gray. Features labeled "> $F_{th}$" appear only above the ablation threshold. **b,c,** Time-resolved transmittance and OPD maps measured below (**b,** 0.9 J cm$^{-2}$) and above (**c,** 1.7 J cm$^{-2}$) the ablation threshold. The pump–probe delay is defined as the temporal separation between the intensity maxima of the pump and probe pulses. T denotes the normalized intensity transmittance relative to the pre-irradiation state. **d,** Temporal traces of transmittance and OPD extracted from the beam center indicated by the black arrow in **b** at −0.13 ps. Colored regions indicate the characteristic temporal regimes discussed in the main text. **e,** OPD map at 0 ps showing a central negative OPD induced by free-carrier excitation and a surrounding weak positive OPD component, inside the black dotted line, attributed to carrier-induced stress. **f,** OPD map at 10 ps showing peripheral negative OPD attributed to long-lived free carriers and surrounding weak positive OPD associated with carrier-induced stress, between the black and white dotted lines. **g,** Expanded nanosecond-scale view of **c** from 8 to

20 ns, showing ablation-plume expansion and the surrounding shock-compressed gas response. **h,** Final-state OPD map showing the ablation crater and a surrounding positive-OPD region associated with permanent material modification. Scale bars, 10 μm.

**Differential phase analysis of early ablating-layer evolution**

The femtosecond-to-millisecond coverage of the present system enables differential analysis between selected delay times within a common imaging framework. By subtracting OPD maps acquired at appropriate delays, slowly varying or quasi-static contributions can be suppressed, thereby isolating transient components associated with specific stages of the ablation process. We first apply this approach to the sub-nanosecond regime, where the transient ablating layer evolves before substantial lateral plume expansion occurs.

Unlike conventional intensity-based measurements, which primarily probe attenuation, scattering, and surface morphology, OPD imaging is sensitive to refractive-index changes along the probe path. This sensitivity allows us to examine optical-property changes associated with the transient ablating layer. Following the temporal framework illustrated in Fig. 3a, we focus on the interval between 0.1 and sub-10 ns, where the solid-to-gas phase transition is expected to have largely completed, while plume expansion above the surface remains limited (22). To isolate rapidly evolving contributions associated with the ablating material, we calculated differential OPD maps between 0.5 and 1.0 ns delay (Fig. 3b). Slowly varying contributions, including lattice-heating-induced refractive-index changes and permanent structural modification, are largely canceled within this narrow temporal window.

Figure 3c shows the resulting differential OPD maps measured below (0.8 and 1.0 J $cm^{-2}$) and above (1.5, 1.9, and 2.3 J $cm^{-2}$) the ablation threshold. Below threshold, where material removal does not occur, the differential signal consists primarily of a ring-shaped positive OPD feature surrounding the excitation region (between black dashed circles). This signal is attributed to relaxation of carrier-induced stress generated during the initial excitation. Analysis of the subsequent temporal evolution confirmed outward propagation consistent with a compressive stress wave in the glass substrate (Supplementary Note 11). Above the ablation threshold, the differential OPD maps exhibit an additional pronounced negative signal localized within the region that later develops into the ablation crater (black arrow). To examine the relation between this early negative phase evolution and the final processed morphology, we compared the differential OPD map with the final-state OPD map (Fig. 3d). The negative differential OPD signal is confined within the final crater region, and the corresponding line profiles show that its lateral extent approximately coincides with the central crater. This spatial correspondence indicates that the material-removal region is already largely established during the 0.5–1.0 ns interval, before substantial lateral plume expansion becomes visible. The negative phase evolution therefore reflects an early optical-property change associated with the evolving ablating material in the future crater region.

We next consider possible mechanisms underlying this negative phase evolution. One possibility is that the material has already transitioned into a plasma state by 0.5 ns and subsequently undergoes axial plasma expansion. In this scenario, the electron density would decrease during expansion, causing the refractive index to recover toward the dielectric value according to the Drude model (Supplementary Note 12). This prediction is opposite to the experimentally observed negative OPD evolution, making simple plasma expansion unlikely to be the sole origin of the measured signal.

Another possible interpretation is that the transient ablating layer remains partially ionized at 0.5 ns and undergoes further ionization or optical-property evolution by 1 ns (Supplementary Note 12). Because the probe beam remains transmissive through the evolving removal region, any plasma contribution must remain below the critical density during this interval. Within a simple Drude description, an increase in free-electron density would drive a negative refractive-index change, consistent with the observed OPD evolution. The signal magnitude is also compatible with this order-of-magnitude picture. At a fluence of 2.3 J cm$^{-2}$, the differential OPD change reaches ~0.2 rad (Fig. 3c, orange arrow). Using the crater depth (~120 nm) as an approximate thickness of the transient ablating layer gives an estimated free-electron density of $1.8 \times 10^{27}$ m$^{-3}$. This value is ~38% of the critical electron density for the 515-nm probe wavelength ($4.8 \times 10^{27}$ m$^{-3}$). Here, the critical density is used only as a reference scale for the Drude-model free-carrier response. This estimate should not be regarded as a unique determination of the free-electron density. Instead, it provides an order-of-magnitude consistency check showing that the measured phase change is compatible with a subcritical transient layer whose optical response continues to evolve before substantial lateral plume expansion. This result illustrates how differential holographic imaging can isolate early removal-stage dynamics that are difficult to access from intensity images alone.

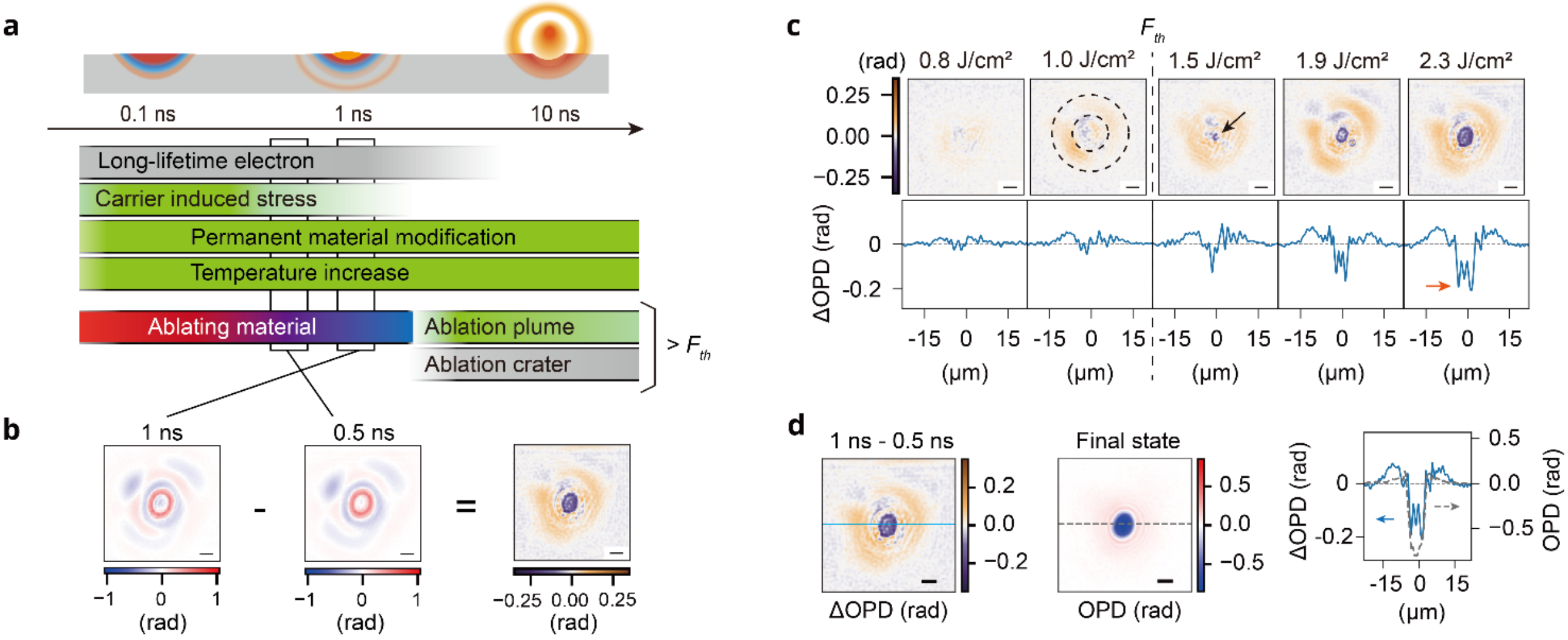

**Figure 3 | Differential phase imaging of early ablating-layer evolution. a,** Schematic timeline of the dominant physical processes contributing to complex-field variations in the sub-nanosecond to nanosecond regime. Differential OPD analysis between selected delays isolates rapidly evolving phase components while suppressing slowly varying background contributions. **b,** OPD maps acquired at pump–probe delays of 0.5 ns and 1.0 ns, together with the corresponding differential OPD map. The differential image highlights transient phase evolution associated with the early removal stage. **c,** Differential OPD maps measured below (0.8 and 1.0 J cm$^{-2}$) and above (1.5, 1.9, and 2.3 J

cm$^{-2}$) the ablation threshold. The lower panels show the corresponding cross-sectional profiles extracted along the x-axis. Above threshold, a negative differential OPD signal appears in the future crater region, indicating continued evolution of the effective optical response of the transient ablating layer before substantial lateral plume expansion. **d,** Comparison between the differential OPD map and the final-state OPD map at 2.3 J cm$^{-2}$. The negative differential OPD signal is confined within the final crater region, indicating that the lateral extent of the material-removal region is already largely established during the 0.5–1.0 ns interval. The right panel shows the corresponding cross-sectional profiles. Scale bar, 5 μm.

**Residual substrate heating and fluence-dependent energy partitioning**

Having established that differential holographic analysis can isolate transient components within selected temporal windows, we next apply it to the central energy-partitioning question: how much of the deposited energy remains as residual heat in the substrate once ablation begins. If additional deposited energy were converted primarily into heat in the remaining substrate, the residual thermal signal would continue to increase with fluence. Conversely, if excess energy were diverted into material removal and ablation-related work, crater growth could proceed without a proportional increase in residual substrate heating. The present measurement provides a direct way to distinguish these scenarios by comparing the final processed morphology with the thermal component remaining immediately after ablation.

To quantify the residual thermal component, we analyzed the substrate response immediately after ablation and before substantial thermal diffusion occurred. Following the temporal framework illustrated in Fig. 4a, residual-heat maps were obtained by subtracting the final-state OPD map, recorded after complete thermal relaxation, from the OPD map acquired at a 1 μs delay. This delay lies well before the characteristic thermal diffusion timescale of ~100 μs, yet after the disappearance of plume-related signatures. The resulting differential OPD therefore isolates the thermal component remaining within the substrate on microsecond timescales after ablation.

Figure 4b shows the resulting residual-heat maps measured below (1.0 J cm$^{-2}$) and above (1.5, 1.9, and 2.3 J cm$^{-2}$) the ablation threshold. These maps represent depth-integrated two-dimensional distributions of residual thermal energy. The corresponding cross-sectional profiles reveal markedly different fluence dependences below and above threshold. Below threshold, the differential OPD increased approximately linearly with fluence, consistent with increased thermal energy deposition at higher excitation fluence. Above threshold, however, the OPD near the beam center reached a plateau despite continued growth of the ablation crater. The gray region in Fig. 4b indicates the above-threshold excitation region in which this saturation behavior was observed.

To quantify this behavior, Fig. 4c compares the center-value differential OPD with the crater depth measured independently by confocal microscopy. Whereas the crater depth increased nearly linearly above the ablation threshold, the residual-heat-associated OPD remained nearly constant. This divergence indicates that increasing excitation fluence enlarges material removal without proportionally increasing the heat retained within the substrate.

As a reference case in which heat diffusion plays a more direct role during energy deposition, we performed corresponding measurements for nanosecond laser ablation. These measurements were analyzed using the same residual-OPD procedure, allowing comparison of the fluence dependence of residual heating between femtosecond and nanosecond excitation. Unlike femtosecond ablation, the residual thermal signal in nanosecond ablation showed no comparable saturation over the investigated fluence range (Fig. 4d and Supplementary Note 13). This contrast supports the interpretation that suppression of residual substrate heating is a characteristic feature of femtosecond laser ablation.

This saturation behavior can be understood from the energy flow specific to femtosecond ablation, as illustrated schematically in Fig. 4e. Below the ablation threshold, the deposited energy remains within the substrate and is eventually converted into heat, leading to an approximately linear increase in the residual thermal OPD with fluence. Above the threshold, however, the energy flow changes qualitatively. Nonlinear attenuation, arising from two-photon absorption followed by free-carrier absorption, sharply reduces the effective penetration depth and confines energy deposition to a thin surface layer (Supplementary Note 7). Once this surface-confined layer undergoes ablation, additional absorbed energy no longer contributes directly to residual heating of the remaining substrate. Instead, it is partitioned into material removal, including phase transitions, ionization, mechanical work, and the kinetic energy of ejected material. As a result, increasing the incident fluence mainly increases the amount of removed material, while the thermal energy left behind in the substrate remains nearly constant. In nanosecond laser processing, by contrast, energy deposition is less strongly confined to a thin surface layer, and heat diffusion proceeds during the pulse. The thermal energy retained in the remaining substrate is therefore expected to increase more continuously with fluence, rather than showing the pronounced saturation observed in the femtosecond case (Fig. 4f). This energy-partitioning mechanism explains the observed saturation of the thermal OPD signal and provides a physical basis for the low-heat-affected nature of femtosecond laser processing.

Finally, we estimated the partitioning of absorbed energy between residual substrate heating and ablation-related energy loss. Assuming that the sub-threshold fluence dependence represents the intrinsic heating response of the substrate, we extrapolated the linear sub-threshold trend to higher fluences (dotted line in Fig. 4c). The difference between this extrapolated OPD and the measured OPD was then attributed to the portion of absorbed energy that would otherwise have contributed to substrate heating but was instead lost through ablation-related processes. Within this simplified framework, the deficit observed at 1.7 J $cm^{-2}$ (black arrow in Fig. 4c) provides an order-of-magnitude estimate suggesting that roughly half of the absorbed energy that would otherwise contribute to residual substrate heating is instead diverted into material removal and ablation-related processes.

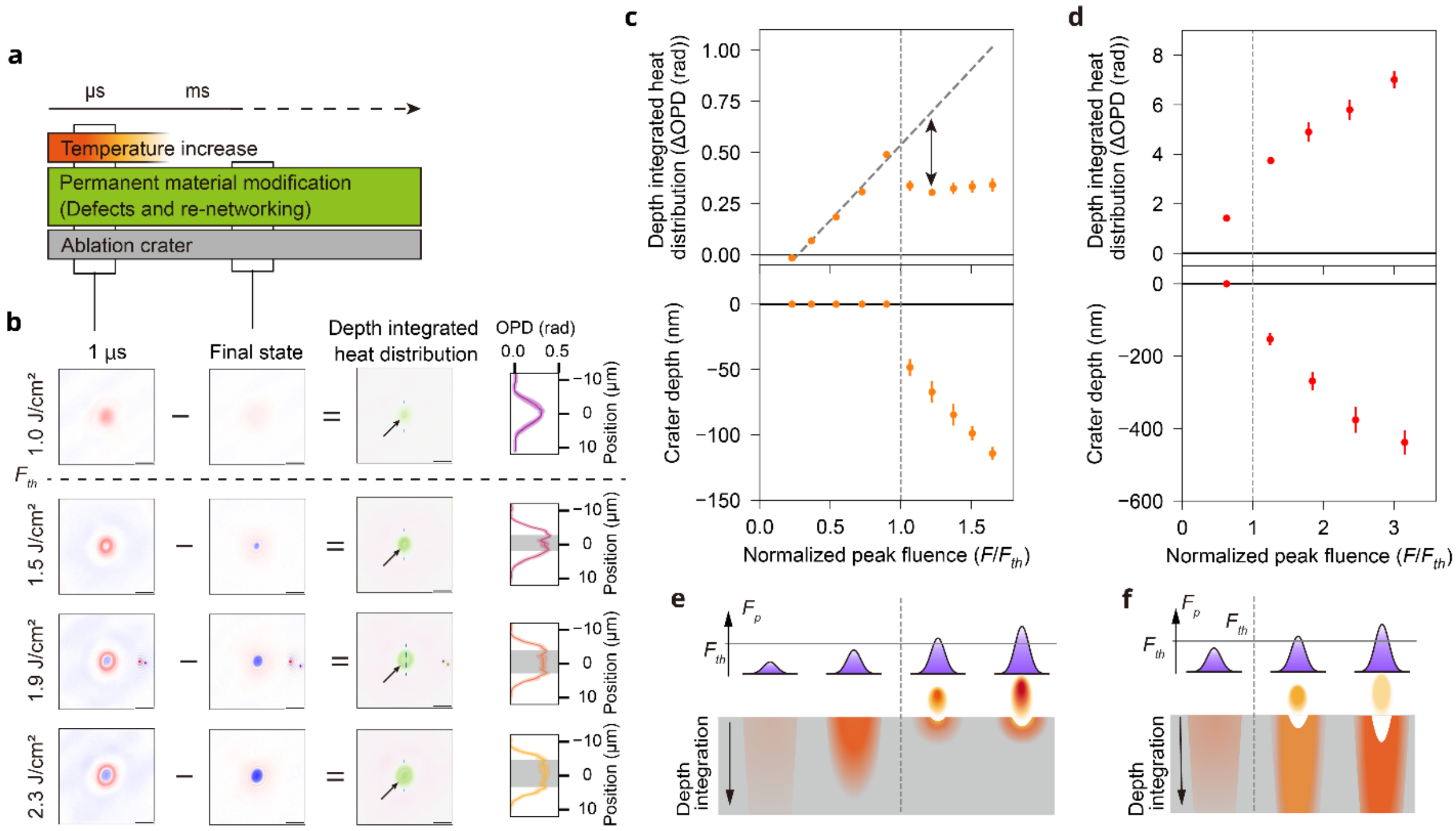

**Figure 4 | Residual substrate heating and fluence-dependent energy partitioning during ablation. a,** Schematic timeline of the dominant physical processes contributing to complex-field variations in the microsecond regime and beyond. Residual heat is isolated by subtracting the final-state OPD map from the OPD map acquired at a 1 μs delay, after plume-related signatures disappear but before substantial thermal diffusion occurs. **b,** OPD maps acquired at 1 μs delay and in the final state, together with the corresponding differential OPD maps that isolate the OPD component associated with residual substrate heating. The rightmost panels show cross-sectional profiles extracted along the dashed lines. The gray region indicates the above-threshold excitation area. Data are shown for fluences below (1.0 J cm$^{-2}$) and above (1.5, 1.9, and 2.3 J cm$^{-2}$) the ablation threshold. **c,** Differential OPD at the pump-beam center, representing the residual heat-associated OPD component, and crater depth measured by confocal microscopy as functions of fluence. The dashed line indicates the extrapolated linear heating trend obtained from the sub-threshold regime. **d,** Corresponding fluence dependences of center-value OPD change and crater depth measured for nanosecond DUV laser processing using the same residual-OPD analysis procedure (Supplementary Note 13). **e,** Schematic illustration of fluence-dependent energy flow during femtosecond laser ablation. Above threshold, additional absorbed energy is preferentially partitioned into material removal and ablation-related processes, leading to saturation of residual substrate heating. **f,** Schematic illustration of fluence-dependent energy flow during nanosecond laser ablation, where residual substrate heating increases more continuously with fluence. Scale bar, 5 μm.

## Discussion

The central contribution of this work is the development of a pump–probe holographic imaging framework that reconstructs laser-ablation dynamics from femtoseconds to milliseconds under matched imaging conditions. By measuring both amplitude and phase, the system tracks optical signatures associated with electronic excitation, stress generation, structural transformation, material removal, plume expansion, permanent modification, and residual heating within a common spatial frame of reference. This capability is important because femtosecond laser ablation is not governed by a single transient state, but by energy flow across electronic, structural, mechanical, and thermal channels that evolve over widely separated timescales.

The broad temporal coverage also enables differential complex-field analysis between selected delay times. In the sub-nanosecond regime, differential OPD imaging reveals a negative phase evolution localized to the future crater region before substantial lateral plume expansion. This response indicates that the effective optical properties of the transient ablating layer continue to evolve during the early removal stage. A simple Drude-based estimate shows that the measured phase change is compatible with a subcritical free-carrier contribution, but the present single-wavelength, depth-integrated measurement does not uniquely determine the ionization dynamics. We therefore interpret this result as phase-sensitive evidence for early ablating-layer evolution, rather than as a definitive measurement of transient free-electron density. More broadly, this analysis illustrates the advantage of complex-field imaging for detecting transient optical responses that are not accessible from intensity measurements alone.

The most direct physical consequence of the femtosecond-to-millisecond measurement is the observation of fluence-dependent energy partitioning above the ablation threshold. By comparing the microsecond OPD component associated with residual substrate heating with the final crater morphology, we find that crater depth continues to increase with fluence while the residual thermal signal retained in the substrate saturates. This divergence shows that additional absorbed energy above the ablation threshold is preferentially redirected into material removal and ablation-related processes rather than being retained as heat in the remaining substrate. The comparison with nanosecond ablation, where no comparable saturation is observed over the investigated fluence range, further supports the interpretation that residual substrate heating is selectively suppressed in femtosecond processing.

This result refines the conventional explanation of the low-heat-affected nature of femtosecond laser processing. The reduced thermal footprint is not only a consequence of limited heat diffusion during the ultrashort pulse. It also reflects a fluence-dependent redistribution of absorbed energy away from residual substrate heating once ablation becomes active. In this sense, low-heat-affected processing is a dynamical energy-partitioning outcome: beyond the ablation threshold, increasing fluence primarily increases the amount of removed material rather than the heat left behind in the substrate.

Several extensions could further increase the physical specificity of this approach. First, the current implementation measures depth-integrated OPD and therefore provides a projected response. Incorporating multi-angle probe illumination within the framework of optical diffraction tomography would enable volumetric reconstruction of transient refractive-index distributions (33), allowing three-dimensional visualization of evolving ablation structures and thermal fields. Second, separation of overlapping processes could be improved by treating time as an additional dimension for multivariate signal decomposition. In the present work, differential analysis isolates components by exploiting temporal differences between physical processes; more advanced temporal unmixing approaches could use the full temporal structure of the dataset, conceptually analogous to spectral unmixing in hyperspectral imaging (34,35). Third, wavelength-resolved probing would introduce refractive-index dispersion as an additional observable and could help distinguish free-carrier or plasma contributions from neutral material responses.

These extensions would be directly relevant to current challenges in ultrafast laser processing. For the early ablating-layer dynamics examined in Fig. 3, wavelength-resolved measurements could test whether the observed phase evolution arises from changes in plasma density, neutral density, structural state, or a combination of these effects. Such information may be particularly important for burst-mode laser processing, where subsequent pulses interact with transient ablation products generated by preceding pulses (36,37). For the residual thermal dynamics shown in Fig. 4, volumetric refractive-index imaging combined with known thermo-optic coefficients would enable reconstruction of transient three-dimensional temperature fields. Quantitative access to heat accumulation and thermal transport could support optimization of multipulse processing strategies, in which cumulative thermal effects strongly influence the final processing outcome (38–40).

Although demonstrated here for femtosecond ablation of BK7 glass, the measurement principle can be broadly applicable to the investigation of laser-induced nonequilibrium dynamics in diverse systems beyond laser processing. Potential applications are, for example, charge-carrier transport in semiconductors, ultrafast photophysical responses in functional materials, and photothermal imaging of living cells (41–44). By extending complex-field imaging from femtoseconds to milliseconds, this approach establishes a general framework for connecting early electronic or molecular excitation with subsequent structural, mechanical, and thermal evolution of matter.

**Conclusion**

We have developed a pump–probe holographic imaging platform that reconstructs amplitude- and phase-resolved optical-field dynamics from femtoseconds to milliseconds under matched imaging conditions. Applied to deep-ultraviolet femtosecond ablation of BK7 glass, this approach connects the successive stages of laser processing, including carrier excitation, stress generation, transient optical and structural changes, material removal, plume expansion, permanent modification, and thermal relaxation, within a single quantitative imaging framework. The broad temporal coverage enables differential phase analysis of both early and late stages of the ablation process. At sub-nanosecond delays, differential OPD imaging reveals a phase evolution localized to the future crater region, indicating that the optical response of the transient ablating layer continues to evolve before substantial lateral plume expansion. At microsecond delays, subtraction of the final-state OPD isolates the residual thermal component retained in the substrate immediately after ablation. The key physical result is the separation of material removal from residual substrate heating across the ablation threshold. Above threshold, the crater depth continues to increase with fluence, whereas the residual thermal signal left in the substrate saturates. This behavior shows that excess absorbed energy is preferentially partitioned into material removal and ablation-related processes rather than retained as residual heat. These results identify fluence-dependent energy partitioning as a dynamical basis for the low-heat-affected nature of femtosecond laser processing and establish femtosecond-to-millisecond holographic imaging as a route to tracking energy flow in laser-driven nonequilibrium material transformations.

## Methods

### Pump-probe holographic imaging system.

Figure S2 shows a detailed schematic of the pump–probe holographic imaging system. A Yb:KGW regenerative amplifier laser system (PHAROS, Light Conversion) generated femtosecond pulses at 1030 nm. These pulses were frequency-converted to generate the femtosecond pump and probe beams as the fourth and second harmonics using a 3.7-mm-thick LBO crystal (EKSMA Optics) and a 0.3-mm-thick BBO crystal (LNG Optics), respectively. The temporal delay between the femtosecond pump and femtosecond probe pulses was controlled using mechanical delay stages (HST-200-GS and HST-50, OptoSigma).

A Yb-doped fiber laser (PYFL-KULT5, KEOPSYS; 1 ns, 1064 nm) served as the nanosecond probe source after frequency doubling in a 30-mm-long LBO crystal (Tecrys). The nanosecond laser operated at 54 kHz, and a 30 Hz subset of pulses was picked using an acousto-optic modulator (1080AFP-D-5.0, G&H) to match the frame rate of the image sensor. Temporal synchronization between the femtosecond pump pulse and the nanosecond probe pulse was controlled using a digital delay generator (DG645, Stanford Research Systems), resulting in a relative timing jitter of 150 ps.

Because femtosecond laser ablation under the present conditions was highly reproducible, the temporal evolution was reconstructed from multiple independent single-pulse measurements. Each pump–probe delay was measured at a fresh sample position rather than by repeatedly probing the same ablation site. For each acquisition, a single DUV pump pulse was focused onto an unprocessed region of the BK7 glass surface. The sample was translated using a motorized stage so that measurements at different pump–probe delays were performed at independent positions, thereby maintaining single-pulse ablation conditions throughout the femtosecond-to-millisecond reconstruction. The pump pulse energy was controlled using a half-wave plate and a reflective polarizer (990-0070-257H, EKSMA Optics), and monitored with a photodetector using leakage light from the dichroic mirror.

The focused pump beam and collimated probe beam were combined coaxially and delivered to the BK7 glass sample. After transmission through a singlet aspheric objective lens (C610TMD-A, Thorlabs) and a relay lens (ACT508-400-A, Thorlabs), the probe beam was directed to a transmission grating (70 line pairs $mm^{-1}$, #46-068, Edmund Optics) positioned at a plane conjugate to the sample. The grating separated the probe beam into the zeroth and first diffraction orders, which served as the object and reference waves for off-axis digital holography. This grating-based lateral-shearing configuration was chosen to provide a common-path reference compatible with broadband femtosecond probe pulses and relatively large processed regions. Common-path holographic configurations offer improved phase stability because the object and reference waves share most of the optical path (45). This configuration avoids possible loss or contamination of low-spatial-frequency components that are important for quantitative transmittance and OPD imaging over a wide field of view. The resulting interferograms were recorded at 30 frames per second using a CMOS image sensor (acA2440-75um, Basler). Post-ablation crater morphologies were characterized using a confocal microscope (VK-X1000, Keyence).

**Complex optical-field reconstruction.**

Figure 1e illustrates the reconstruction procedure for retrieving the complex optical field from a recorded hologram. Each hologram was first transformed into the spatial-frequency domain by two-dimensional Fourier transformation, yielding two interferometric sidebands and a central non-interferometric component. One of the sidebands was isolated using a spatial-frequency cropping window corresponding to an effective reconstruction numerical aperture of approximately 0.4. An inverse Fourier transform of the selected sideband then reconstructed the complex optical field of the probe beam.

Intensity transmittance and OPD maps were calculated from the reconstructed complex field. Unless otherwise stated, these maps presented in this work are expressed as differential signals relative to the pre-irradiation state acquired before pump-pulse excitation. This procedure removes static background contributions from the sample and optical system and enables quantitative comparison of transient complex-field responses across different pump–probe delays.

**Aberration compensation and digital focusing.**

Spherical aberration introduced by the singlet aspheric objective lens was corrected numerically. The objective lacked a correction collar, whereas conventional collar-corrected objectives were unsuitable because the high-intensity probe pulses caused optical damage within the objective optics near the Fourier plane. Aberrations originating from the glass–air interface were compensated digitally by applying the inverse aberration-formation process using angular-spectrum propagation. After aberration correction, precise digital refocusing was performed using the Tamura-of-the-gradient (ToG) focus criterion (46). Additional methodological details are provided in Supplementary Note 4. For each dataset, the digital refocusing distance was determined from the post-processed image, where the transmittance image became closest to transparent and the crater rim in the OPD image was most clearly resolved. The same propagation distance was then applied to all pump–probe delays and was not independently optimized for each time-resolved image.

**Image registration and differential OPD analysis**

For differential analysis between pump–probe delays, reconstructed OPD maps were registered onto a common pixel grid before subtraction. This registration was applied to minimize apparent OPD differences arising from positional shifts between independent single-pulse measurements at different sample locations. Differential OPD maps were then calculated between selected delay times to isolate specific components of the laser-ablation dynamics. For the sub-nanosecond analysis in Fig. 3, OPD maps acquired at 0.5 and 1.0 ns delays were subtracted to suppress slowly varying contributions and highlight rapidly evolving phase components associated with the early removal stage. For the residual-heating analysis in Fig. 4, the final-state OPD map, recorded after complete thermal relaxation, was subtracted from the OPD map acquired at a 1 μs delay. This subtraction isolates the OPD component associated with residual substrate heating immediately after ablation and before substantial thermal diffusion.

**Fluence-dependent analysis of residual heating and crater depth**

For the fluence-dependent measurements shown in Figs. 4c and 4d, each data point was obtained from five independent single-pulse measurements performed at fresh sample positions under nominally identical irradiation conditions. The OPD component associated with residual substrate heating was evaluated as the differential OPD between the image acquired at a 1 μs delay and the final-state image, using the same reconstruction, registration, and subtraction procedures for all repetitions. The crater depth was measured independently by confocal microscopy after irradiation. The plotted values represent the mean of the five measurements, and the error bars indicate the standard deviation.

For the approximate energy-partitioning estimate in Fig. 4c, the linear fluence dependence observed below the ablation threshold was used as a reference for the intrinsic heating response of the substrate. This sub-threshold trend was extrapolated to higher fluences, and the difference between the extrapolated OPD and the measured OPD was attributed to the portion of absorbed energy that would otherwise have contributed to residual substrate heating but was instead diverted into material removal and ablation-related processes. This estimate is intended as an order-of-magnitude analysis and does not account for possible fluence-dependent changes in absorption, reflection, plasma shielding, or the detailed thermodynamic pathways of material removal.

**Acknowledgments**

This work was financially supported by the MEXT Quantum Leap Flagship Program (MEXT Q-LEAP) (JPMXS0118067246), the Cross-ministerial Strategic Innovation Promotion Program (SIP), "Photonics and Quantum Technology for Society 5.0", JST FOREST Program (JPMJFR236C), and RIKEN TRIP initiative. S.K. was supported by the University of Tokyo MERIT-WINGS Program.

**Author Contributions**

S.K., M.T. and H.S. developed the system. S.K. performed the experiments and analyzed the data. S.K., K.T., H.S., K.K. and T.I. discussed the interpretation of the results. T.I. supervised the work. S.K., K.T., and T.I. wrote the manuscript with inputs from other authors.

**Data, Materials, and Software Availability**

The data provided in the manuscript is available from the corresponding author upon reasonable request.